\documentclass[12pt]{article}
\usepackage{amsmath, amssymb, graphics, setspace}
\usepackage{slashbox}
\usepackage{graphicx}
\begin{document}
\begin{center}

{\large\bf Theoretical basis for the unification of the integer and the fractional quantum Hall effects  }
\vskip 1cm
 Shuming Long $^{a}$, Jianhua Wang$^{a}$, Kang Li $^{b}$\footnote{Corresponding email: kangli@hznu.edu.cn} and Yi Yuan$^{a}$
\\\vskip 1cm

{\it\small a.~~ Shaanxi University  of Technology, Hanzhong, 723001,
Shaanxi, China

b.~~ Hangzhou Normal University, Hangzhou, 310036, Zhejiang, China
\\}
\end{center}

\begin{abstract}
\noindent This paper intends to provide a theoretical basis for the unification of the integer and the fractional quantum Hall effects. Guided by concepts and theories of quantum mechanics and with the solution of the Pauli equation in a magnetic field under the symmetric gauge, wave functions, energy levels of single electrons, and the expectation value of electron's spatial scope are presented. After the quotation of non-interaction dilute gas system, the product of single electron's wave functions is used to construct wave functions of the N electron gas system in magnetic field. Then the expectation value of the system's motion area and the electron's surface density are obtained. In this way, the unification explaination of the integer and the fractional quantum Hall effects is formulated without the help of the concept of fractional charge.


\noindent PACS number(s): 72.20.My, 71.45.-d, 73.40.Lq
\end{abstract}

\section{Introduction}
In recent years the Hall effect attracts increasing attention. In fact,the Hall effect was first introduced in 1879 by Edwin Hall in an experiment which showed when a conductor carrying an electric current perpendicular to an applied magnetic field develops
a voltage gradient which is  transverse to both the current and the
magnetic field\cite{klizing1}. About 100 years later, quantum Hall effect was discovered by  Klaus von Klitzing, when he was studying conductive properties of semiconductors in an very low temperature and strong magnetic field. Because of this, he won the Nobel Prize in physics in 1985\cite{klizing2}. After 1982, under a stronger magnetic field and  lower temperature, Daniel Chee Tsui, Horst L. Stormer and Robert B.Laughlin found fractional quantum Hall effect and won the 1998 Nobel Prize in physics\cite{tsui1}-\cite{laughlin1}.

The quantum Hall effects have remained one of the most spectacular discoveries in physics in the past century. With two dimensional electron gas of the semiconductor which affected by a strong magnetic field perpendicular to the surface, Hall resistance  $\rho_{xx} $ and  $\rho_{xy}$  will have a change under the changing of the surface density of electron gas, and the points  $\rho_{xx}=0 $ will appear periodically. Near these points, $\rho_{xy}$ remains the same in a certain changing range of the density. This is the Hall resistance's plateau phenomena, it is also called Hall plateau in short. On Hall plateau, resistance $\rho_{xy}=h/(\nu e^2)$ £¬in which $\nu$ is an integer. When $\nu$ is a fraction, the Hall plateau phenomena is known as fractional quantum Hall effect, it was also found in the experiment \cite{tsui1}-\cite{ASW}.

However, some demerits can be found in previous studies on the effect. For instance, the Pauli equation of charged particles was solved in Landau gauge so as to describe the integral Hall effect. Such kind of method is not convincing in theory because it loses some important information by degrading an axial-symmetric problem to a problem on one-dimensional harmonic oscillator, resulting in anthropogenic quantization of momentum of free motion. Besides this example, symmetric gauge was also used by R.B.Lauphline to solve charged particles' Pauli equation. With the wave functions of single electrons, the expectation value of electron motion's area in a ground state was calculated before wave functions of the electron gas system were obtained. In this method fractional charge were used to explain fractional quantum Hall effect.

In this paper, with the accurate solution of the Pauli equations, energy levels and wave functions of single electron in a magnetic field are first presented. Then, the probability density of a moving single electron and its energy degeneracy per unit area are calculated. These results can be used in a two-dimensional non-interaction electron gas system and obtain the formula of quantum Hall resistance which can explain both integral Hall effect and fractional Hall effect. It is concluded that energy degeneracy is theoretically infinite when electron orbital magnetic moment has the same direction as external magnetic field (i.e.$m\leq0$). This is supposed to be the very cause of the quantum Hall effect. In next section, the pauli equation is solved, wave function and energy levels for single electron are given explicitly. In section 3, the probability density distribution function in a strong magnetic field is obtained. The degeneracy density is studied in section 4, finally, some remarks and conclusions are given in the last section.

\section{ Energy levels and wave functions of an electron moving in a magnetic field }
When an electron moves in the magnetic field, it should be described by Dirac equation as follows,
\begin{equation}
i\hbar \frac{\partial }{\partial t}\psi =H\psi , H=c\overset{\to }{\alpha }.\left(\frac{\hbar }{i}\triangledown +\frac{e}{c}\overset{\to }{A}\right)-e \varphi +\mu c^2\beta
\end{equation}
In terms of non-relativistic limit of Dirac equation, eliminating $\text{$\mu $c}^2$ and ignoring the term of electric field effect $ \text{\textit{ie}}\text{$\hbar $c$\alpha $}.\overset{\to }{\text{\textit{$E$}}}$, we have the following Schrodinger equation,
\begin{equation}
i\hbar \frac{\partial }{\partial t}\psi =[\frac{1}{2\mu }\left(\frac{\hbar }{i}\triangledown +\frac{e}{c}\overset{\to }{A}\right)^2-e \varphi +\frac{e\hbar }{2\mu c}\sigma .B]\psi
\end{equation}
in which $\text{e$\hbar $}/(2\text{$\mu $c})\text{\textit{$=$}}9.274 10^{-21}\text{\textit{$(\text{Erg}/\text{Gauss})$}}$ called Bohr magneton which is a close to a magnetic moment of an electron. Taking two dimensional spin matrix operator as Pauli matrix,  Eq. (2) can describe electron motion at a low speed in an external electromagnetic field.
If an electron with a mass $ \mu $ goes at an initial momentum $p_z$ along $ z$-direction  into the magnetic field of magnetic vector potential $\overset{\to }{A}=(-\text{\textit{$By$}}/2, \text{\textit{$Bx$}}/2 , 0)$( a symmetric gauge here is a natural physical selection), the curl of vector potential A gives magnetic induction strength $\text{\textit{$\text{\textit{$($}}\text{\textit{$B$}}_x,B_y,0\text{\textit{$)$}}=(0,0,B)$}}$. Since the external electric field does not affect electrons' probability distribution but the system's energy levels by global translation, so the electric field effect can be ignored, thus Schrodinger equation above is simplified to the Pauli equation,

\begin{equation}
i\hbar \frac{\partial }{\partial t}\psi =[\frac{1}{2\mu }(\frac{\hbar }{i}\frac{\partial }{\partial x}-\frac{eB}{2c}y)^2+\frac{1}{2\mu }(\frac{\hbar }{i}\frac{\partial }{\partial y}+\frac{eB}{2c}x)^2+\frac{- \hbar ^2}{2\mu }\frac{\partial ^2}{\partial z^2}+\frac{e \text{$\hbar $B}}{2\mu c}\sigma _z]\psi
\end{equation}
Since electrons in the z direction is  free  and the periods of an angular motion,  we separate z direction and angular motion from spatial and temporal variation, the wave function reads
\begin{equation}
\psi (\rho ,\varphi \, ,\, z,\, s\, ,\, t\, )=\phi (\rho ,\, s)e^{i\, m\, \varphi }e^{-i\, p_z\, z/\hbar }e^{-i\, E\, t/\hbar }, m=0,\pm 1,\pm 2,\text{...}
\end{equation}
where $s$ is the electron's spin state variable. The function $\phi (\rho ,\, s)$, in cylindrical coordinates, satisfies,
\begin{equation}
[-(\frac{\partial ^2}{\partial \rho ^2}+\frac{1}{\rho }\frac{\partial }{\partial \rho }-\frac{m^2}{\rho ^2})+(\frac{eB}{2\hbar c})^2\rho ^2+\frac{eB}{2\hbar c}(2m+2\sigma _z)]\phi (\rho ,s)=\frac{2\mu E-p_z{}^2}{\hbar ^2}\phi (\rho ,s)
\end{equation}
After the separation of variables on radial motion and spin motion, Eq. (5) becomes
\begin{equation}
\phi (\rho ,s)=u(\xi )\chi _{\lambda }(s),\, \, \xi =\rho /a,\, \, a=(\frac{2\hbar c}{eB})^{1/2}
\end{equation}

\begin{equation}
\sigma _z\chi _{\lambda }(s)=\lambda \chi _{\lambda }(s),\, \, \lambda =1,\, \chi _{\lambda }(s)=\chi _+(s)=\left(
\begin{array}{c}
 1 \\
 0
\end{array}
\right), \lambda =-1,\, \chi _{\lambda }(s)=\chi _-(s)=\left(
\begin{array}{c}
 0 \\
 1
\end{array}
\right)
\end{equation}

\begin{equation}
[-(\frac{\partial ^2}{\partial \xi ^2}+\frac{1}{\xi }\frac{\partial }{\partial \xi }-\frac{m^2}{\xi ^2})+\xi ^2]u(\xi )=Ku(\xi ),\, \, K=a^2\frac{2\mu E-p_z{}^2}{\hbar ^2}-2(m+\lambda )
\end{equation}
With $u(\xi )=w(\xi )\exp (-\xi^2/2)$£¬ Eq. (8) changes to
\begin{equation}
-w''(\xi )+(2\xi -\frac{1}{\xi })w'(\xi )+(2-K+\frac{m^2}{\xi^2})w'(\xi )=0,\, \, K=a^2\frac{2\mu E-p_z{}^2}{\hbar ^2}-2(m+\lambda )
\end{equation}
With the method of the power series expansion \cite{Long}, the solutions of Eq. (8) are
\begin{equation}
u_1(\xi )=e^{- \frac{\xi ^2}{2}}\sum _{k=0}^{\infty } \frac{(\frac{2+2m-K}{4})_k}{k!(1+m)_k}\xi ^{2k+m} , u_2(\xi )=e^{- \frac{\xi ^2}{2}}\sum _{k=0}^{\infty } \frac{(\frac{2-2m-K}{4})_k}{k!(1-m)_k}\xi ^{2k-m}
\end{equation}
In Eq. (10) $(m)_k=m(m +1)\ldots (m +k-1)$ is called Pochhammer symbol.
It is known that the wave function is required to be bounded at $\xi =\rho /\text{\textit{$a$}}=\text{\textit{$0$}}$. Since quantum number $\text{\textit{$m$}}$ can be positive or negative, $\text{\textit{$m$}}$ in the first solution of Eq.(10) can be equal to or more than 0, $\text{\textit{$m$}}$ in the second solution of it can be only negative integers. Therefore, Eq.(8)'s meaningful solution can only be
\begin{equation}
u(\xi )=\xi ^{|m|}\, e^{\left.-\xi ^2\right/2}\sum _{k=0}^{\infty } \frac{(\frac{2+2|m|-K}{4})_k}{k!\, (k+|m|\, )!}\, \xi ^{2k}
\end{equation}
Bound state wave functions are required to tend to zero as soon as possible when space variables tend to infinity.  The solution of Eq.(11) is concord to function $ exp(\xi^2 /2)$. Thus, parameter $(2+2|m|-K )/4 $ in the equation must be equal to negative integers denoted by $-n_\rho$ ,then for any value of $m$, the solution reduce to polynomial. So the parameter K is
\begin{equation}
K=4n_{\rho }+2|m|+2=a^2\frac{2\mu E-p_z{}^2}{\hbar ^2}-2(m+\lambda ),m,\, n_{\rho }=0,1,2,\text{...}
\end{equation}
 we get the meaningful solution and energy eigenvalue of Eq.(8) respectively,
 \begin{equation}\label{wavefunction}
    u_{n_\rho |m|}(\xi)=N_{n_\rho |m|}\xi^{|m|}e^{-\xi^2/2}\sum_{k=0}^{n_\rho}\frac{(-1)^k{\rm Binomial}(n_\rho,k)}{(k+|m|)!}\xi^{2k},
 \end{equation}

\begin{equation}
E_{n_{\rho }m\, \lambda }=\frac{p_z{}^2}{2\mu }+(n_{\rho }+\frac{|m|+m+\lambda +1}{2})\frac{e\hbar B}{\mu c}, \text{\textit{$ $}}\text{\textit{$m$}}\text{\textit{$,$}}\text{\textit{$\, $}}\text{\textit{$n_{\rho }$}}=0,1,2,\text{...}
\end{equation}
in which $\lambda$ can take $1$ or $-1$, and Binomial(n,m) in Eq.(13) is a total number for taking $ m$ out of $n$. In Eq.(14)  in which $e\hbar B/ (\mu c)=2\mu_B B=2\hbar\omega_L$,  $\mu_B=e\hbar/(2\mu c)$   is a Bohr magneton,  and $\omega_L=eB/(2\mu c)$ is Larmor rotation angle frequency. In an external magnetic field an electron's energy is divided into three parts. The first part is  translational kinetic energy  produced by the electron moving along $z$ direction; the second one is   produced by the electron's radial motion; and the third part is the interaction energy   of the electron's total magnetic moment and the external magnetic field.
The wave function satisfying Eq.(3) is
\begin{equation}
\psi _{n_{\rho }m\, \lambda }(\rho ,\varphi \, ,\, z,\, s\, ,\, t\, )=u_{n_{\rho }|m|}(\rho /a)\chi _{\lambda }(s)e^{i\, m\, \varphi }e^{-i\, p_z\, z/\hbar }e^{-i\, E\, t/\hbar }\,
\end{equation}
in which, $n_\rho=0,1,2,\ldots; m=0,\pm 1, \pm 2, \ldots $ .  Using the normalization condition, we obtain the normalization coefficient of Eq.(13),
\begin{equation}
N_{n_{\rho }|m|}=(\frac{(n_{\rho }+|m|)!}{a^2\pi \, n_{\rho }!}){}^{1/2},\text{   }a=(\frac{2\hbar c}{eB})^{1/2}
\end{equation}

\section{The electron's probability density distribution under a strong magnetic field}

With a Light speed of $c=2.998 10^{10}$(cm/sec), an electron charge of $e=4.803\ 10^{-10}$(Statcoulomb), Planck constant $\hbar \text{\textit{$=$}}1.055 10^{-27}$(Erg sec),  electron mass$\mu =9.109\ 10^{-28}$(gram), an external magnetic field B=150000(Gauss), and the electron's speed along Z direction is $v=3\times 10^5$ cm/sec,  we get of $\left.p_z{}^2\right/(2\mu )=2.562\ 10^{-5}\text{eV},$ $\varepsilon _0=\mu _BB=1.739\ 10^{-3}\text{eV},\text{\textit{$a$}}=93.68\overset{o}{A}.$
Above calculation shows that electron motion in a magnetic field with B=150000 gauss has small energy and moves in a "big" space. The interaction energy of the electron and magnetic field is as small as per 1.74 electron-volt. Compared with this energy, electronic low speed ($v\leq$1000m/sec) average kinetic energy is much smaller, less than three hundred thousandths electron volts. Under the restriction of the magnetic field in the ground state electronic space motion range scale is 177 times of that of hydrogen atoms. So, electrons in a magnetic field is more free then in an atom. And electron's energy under the constraint of a magnetic field is four grade smaller than that under the constraint of the nucleus coulomb field in an atom. In this way can dilute electron gas in a magnetic field be regarded as non-interaction electron gas system.
The probability density function from Eq.s (13) -(16) is related to electron motion's radial coordinate position, radial quantum number $n_\rho$  and angular quantum number $|m|$. In a cylindrical shell with a radius from $\rho$  to $\rho+ d\rho $,  the probability $ dP$ is,

\begin{equation}
dP=f_{n_{\rho }m\, }(\rho )d\rho ,\text{   }f_{n_{\rho }m\, }(\rho )=2\pi \rho \left|u_{n_{\rho }|m|}(\rho /a)|^2\right.
\end{equation}
In a magnetic field the radial probability density distributions of electron $f_{n_\rho m}(\rho)$  are shown in Figure I, which corresponds to $n_{\rho }=(0,4,8,12), m=(0,4,8)$, where we take $ B=150000 $ Gauss, and  spin quantum number$\lambda=1$.

\begin{figure}
\includegraphics[width=5.6cm,height=4.2cm]{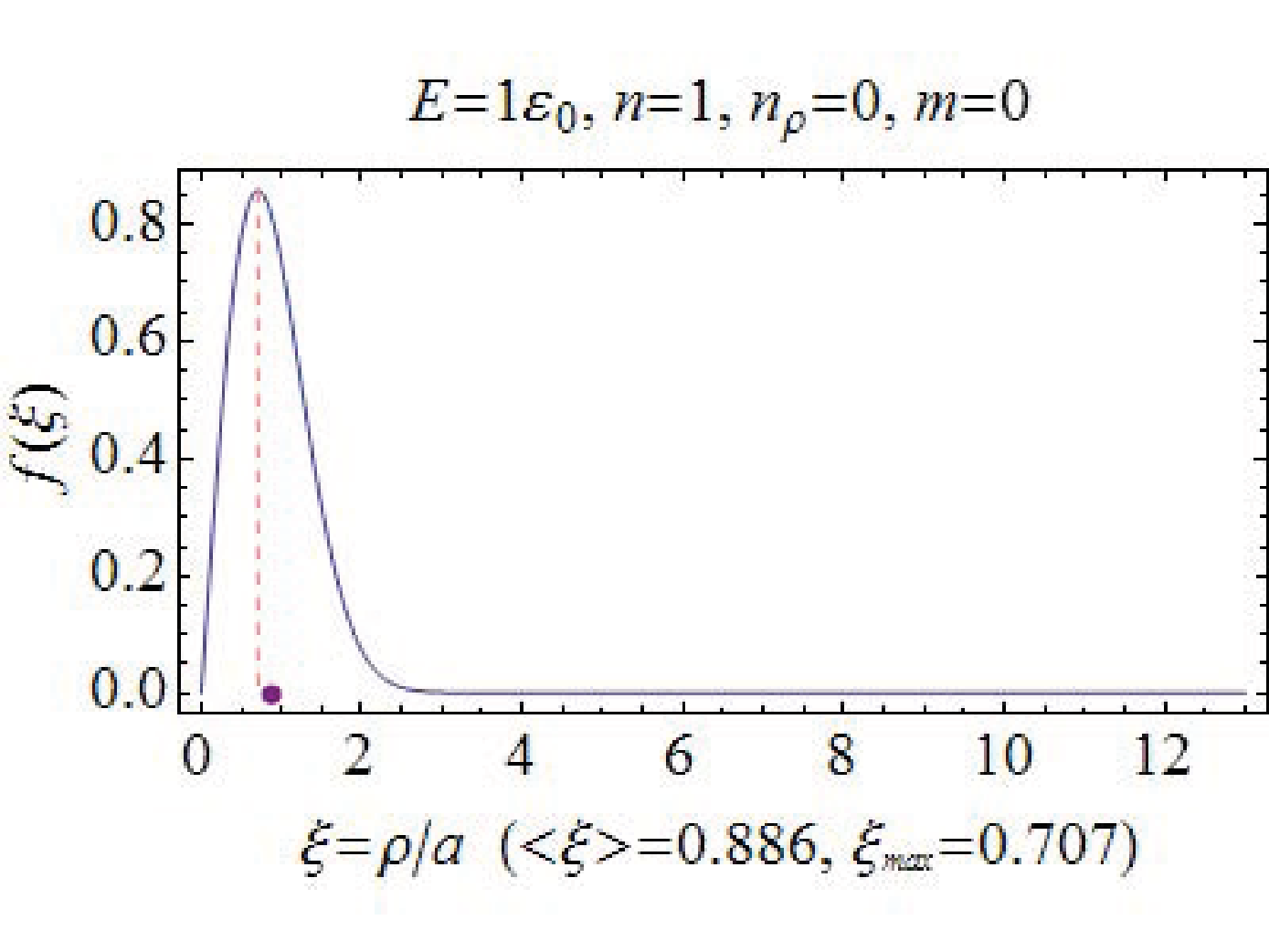}
\includegraphics[width=5.6cm,height=4.2cm]{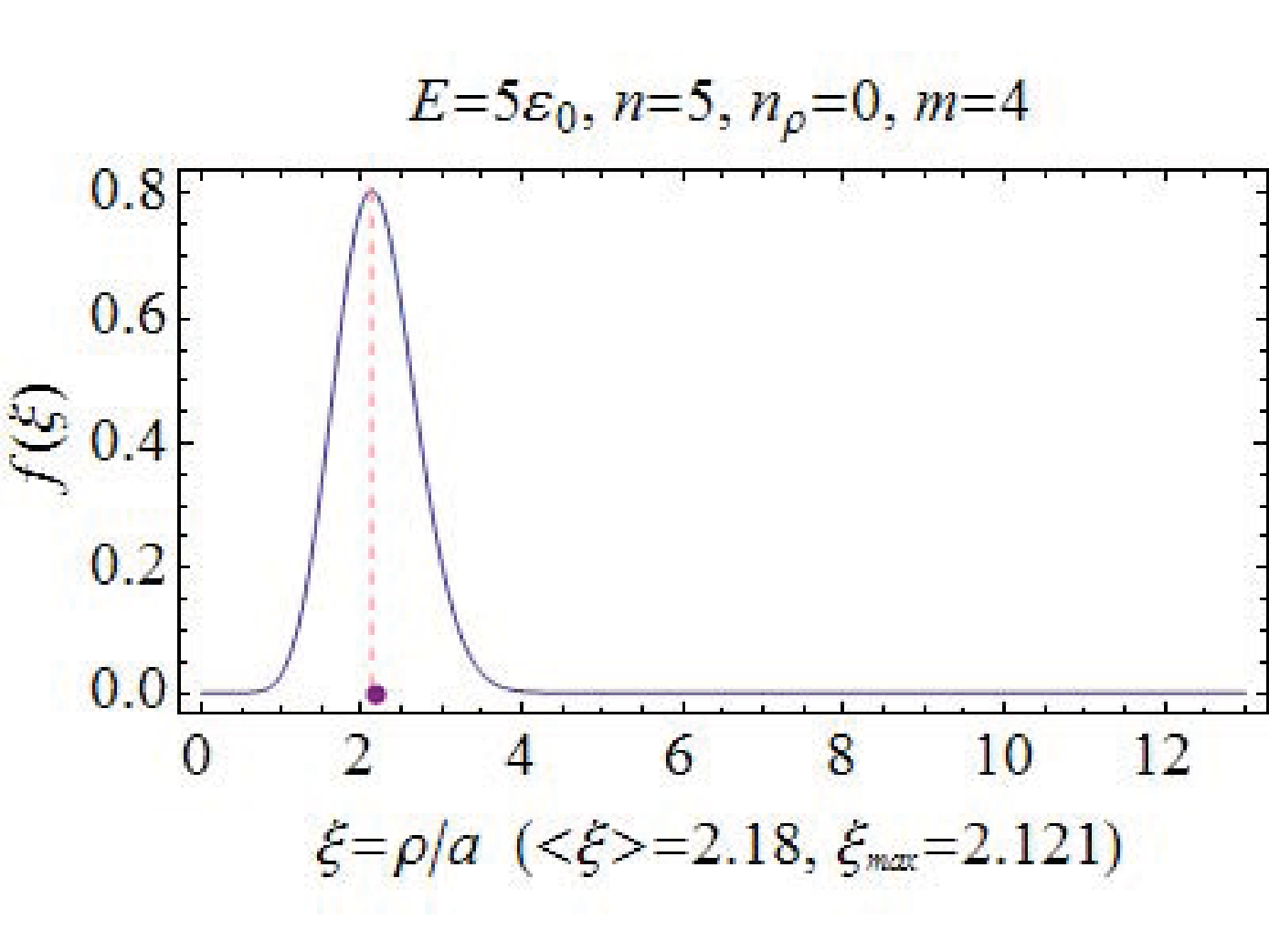}
\includegraphics[width=5.6cm,height=4.2cm]{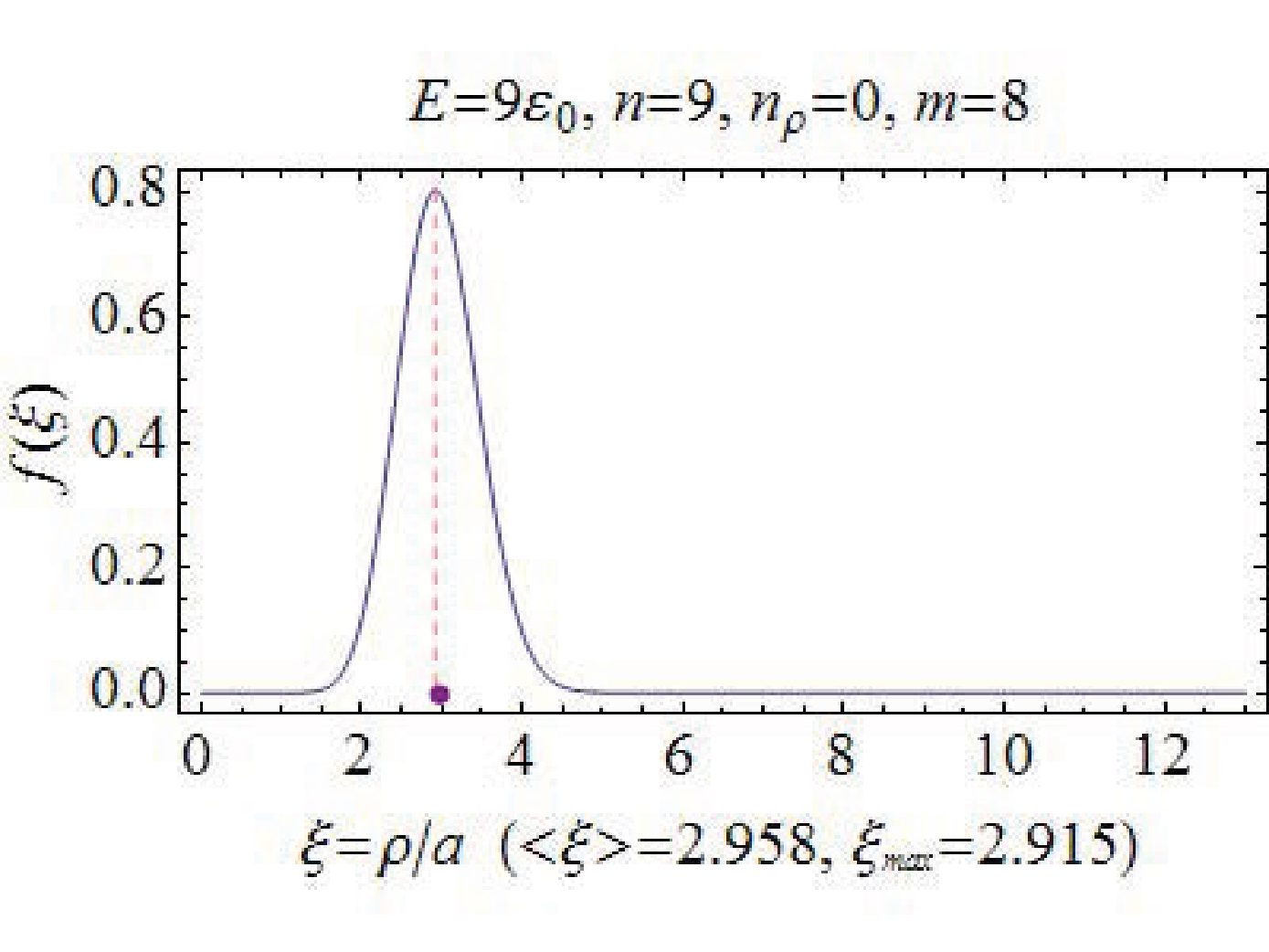}\\
\includegraphics[width=5.6cm,height=4.2cm]{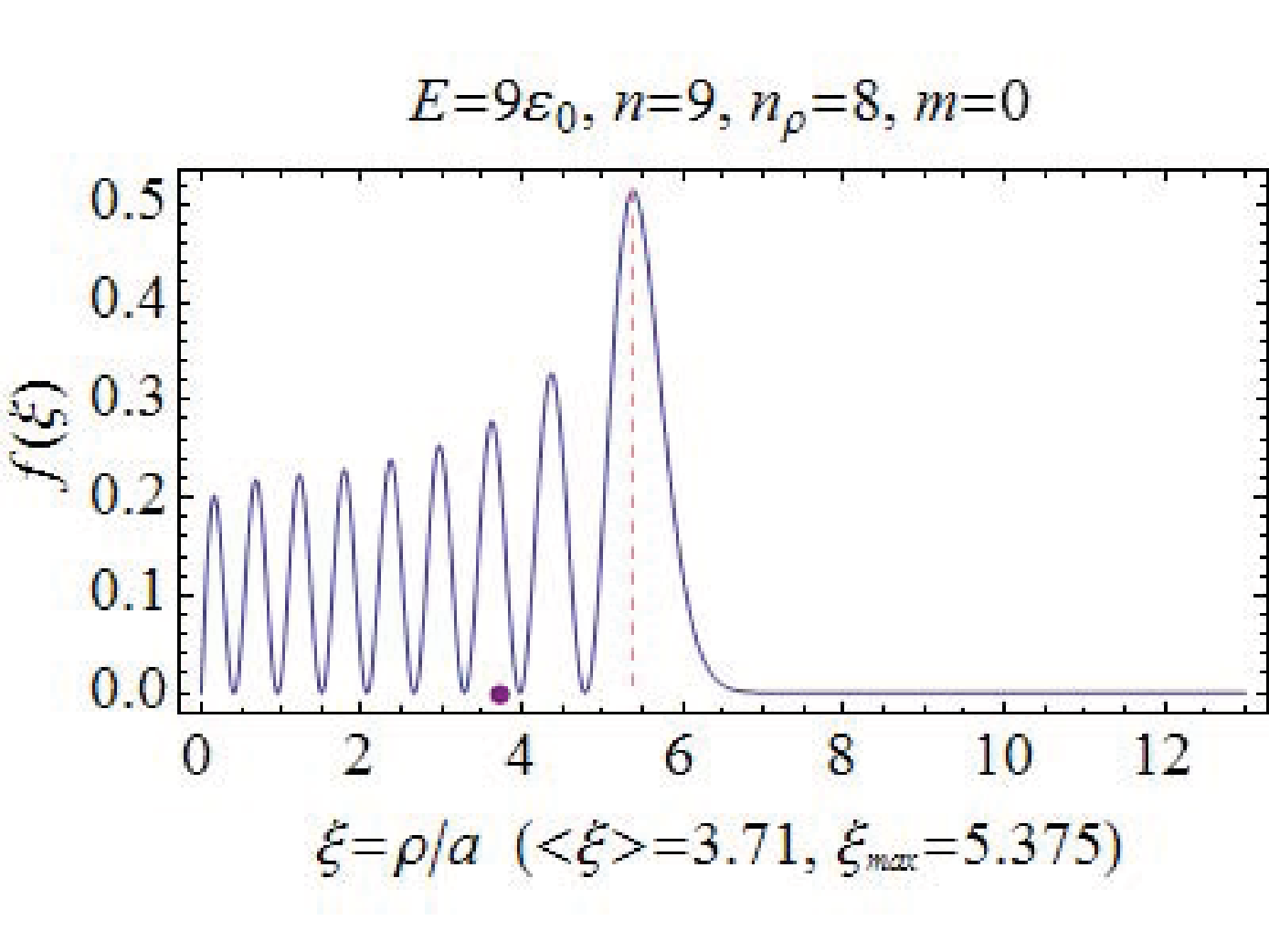}
\includegraphics[width=5.6cm,height=4.2cm]{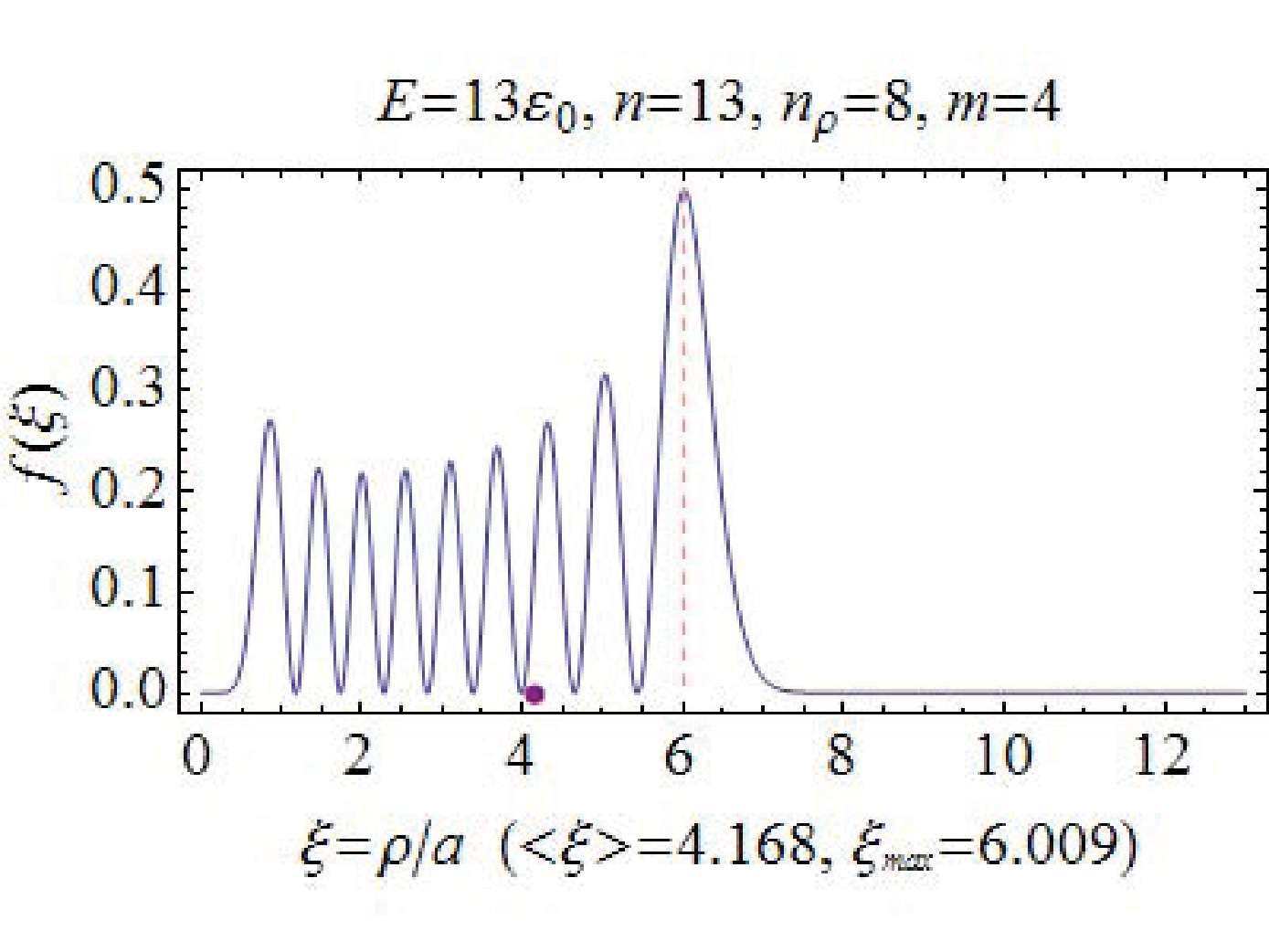}
\includegraphics[width=5.6cm,height=4.2cm]{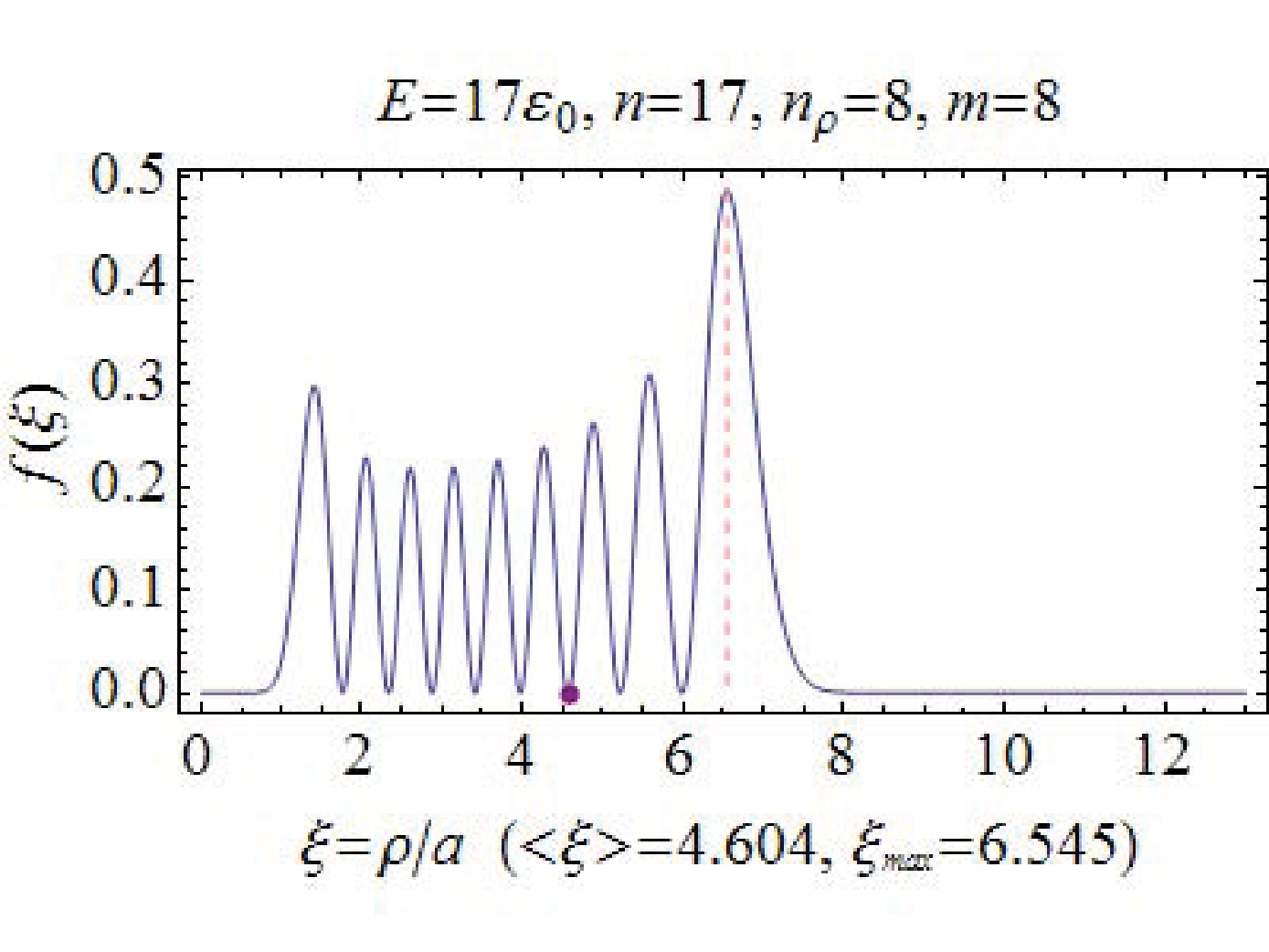}\\
\includegraphics[width=5.6cm,height=4.2cm]{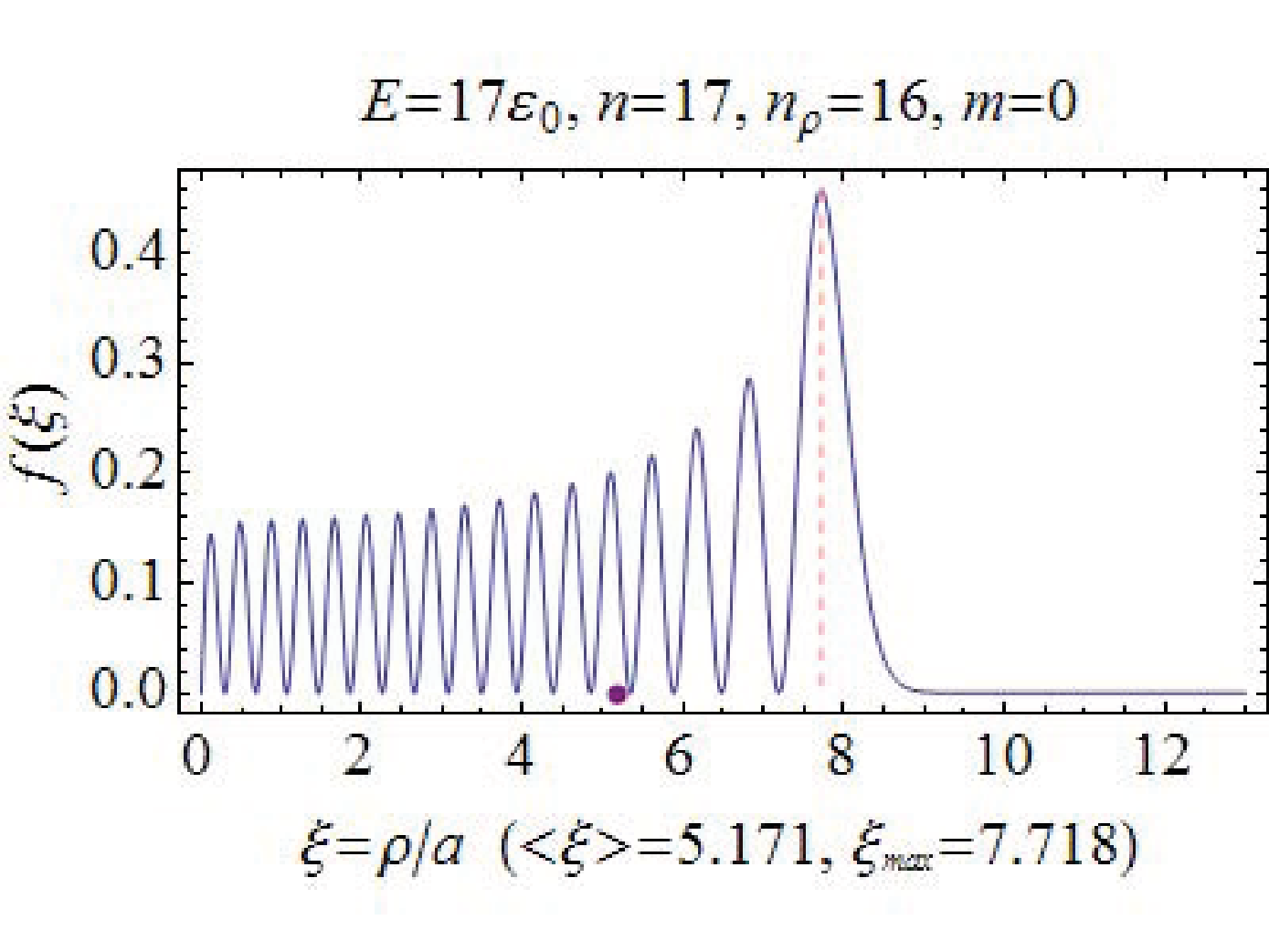}
\includegraphics[width=5.6cm,height=4.2cm]{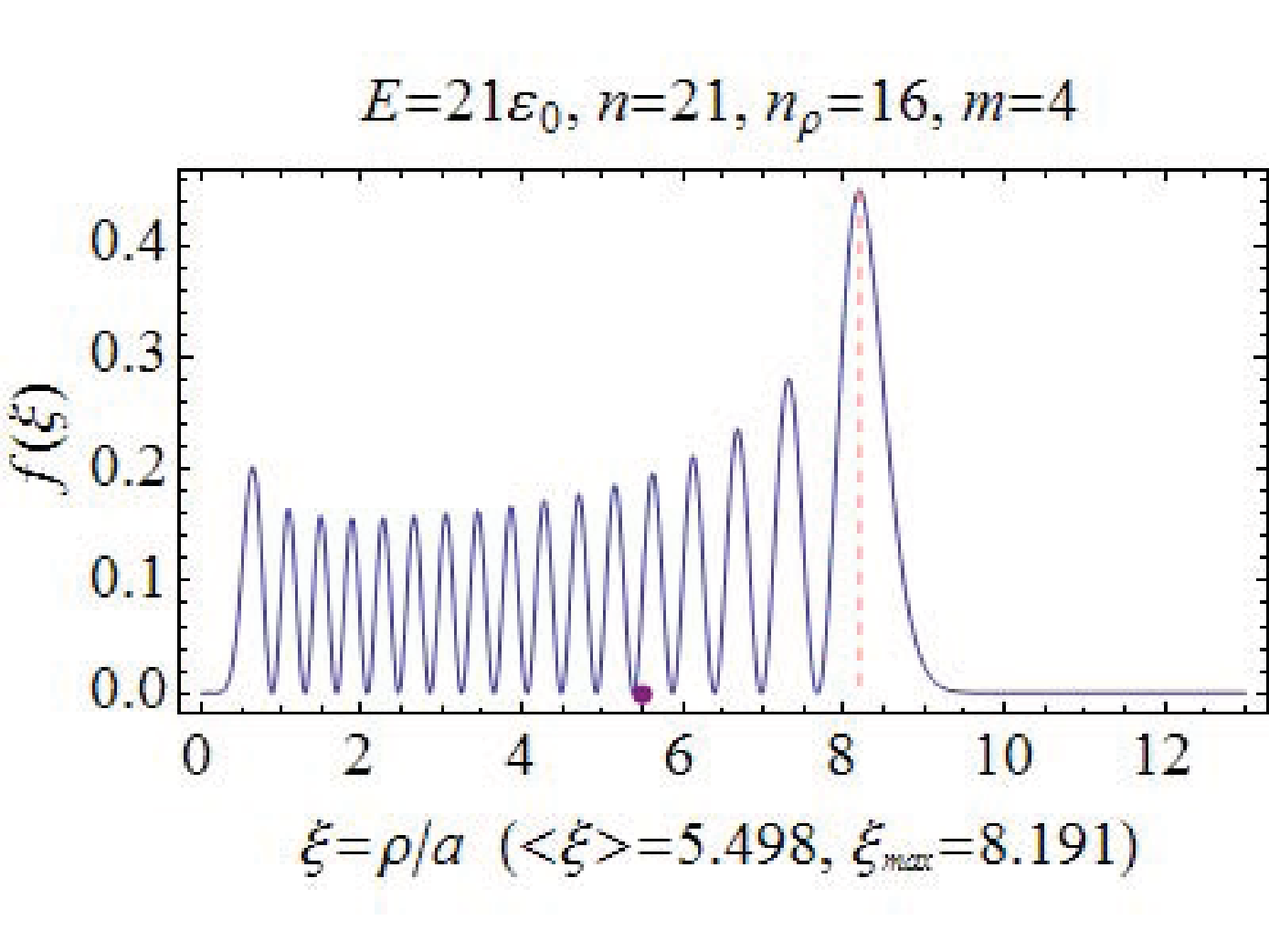}
\includegraphics[width=5.6cm,height=4.2cm]{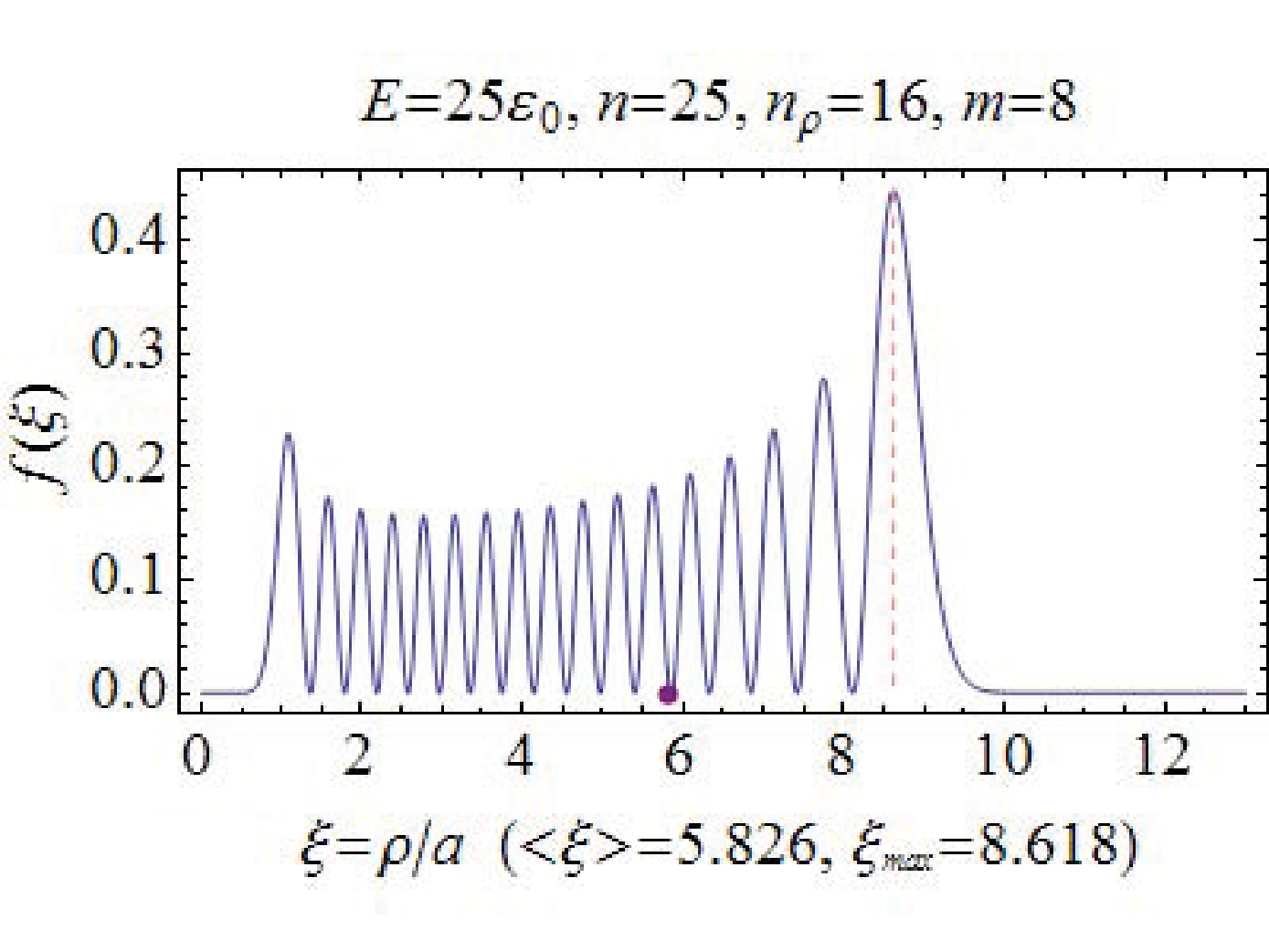}\\
\includegraphics[width=5.6cm,height=4.2cm]{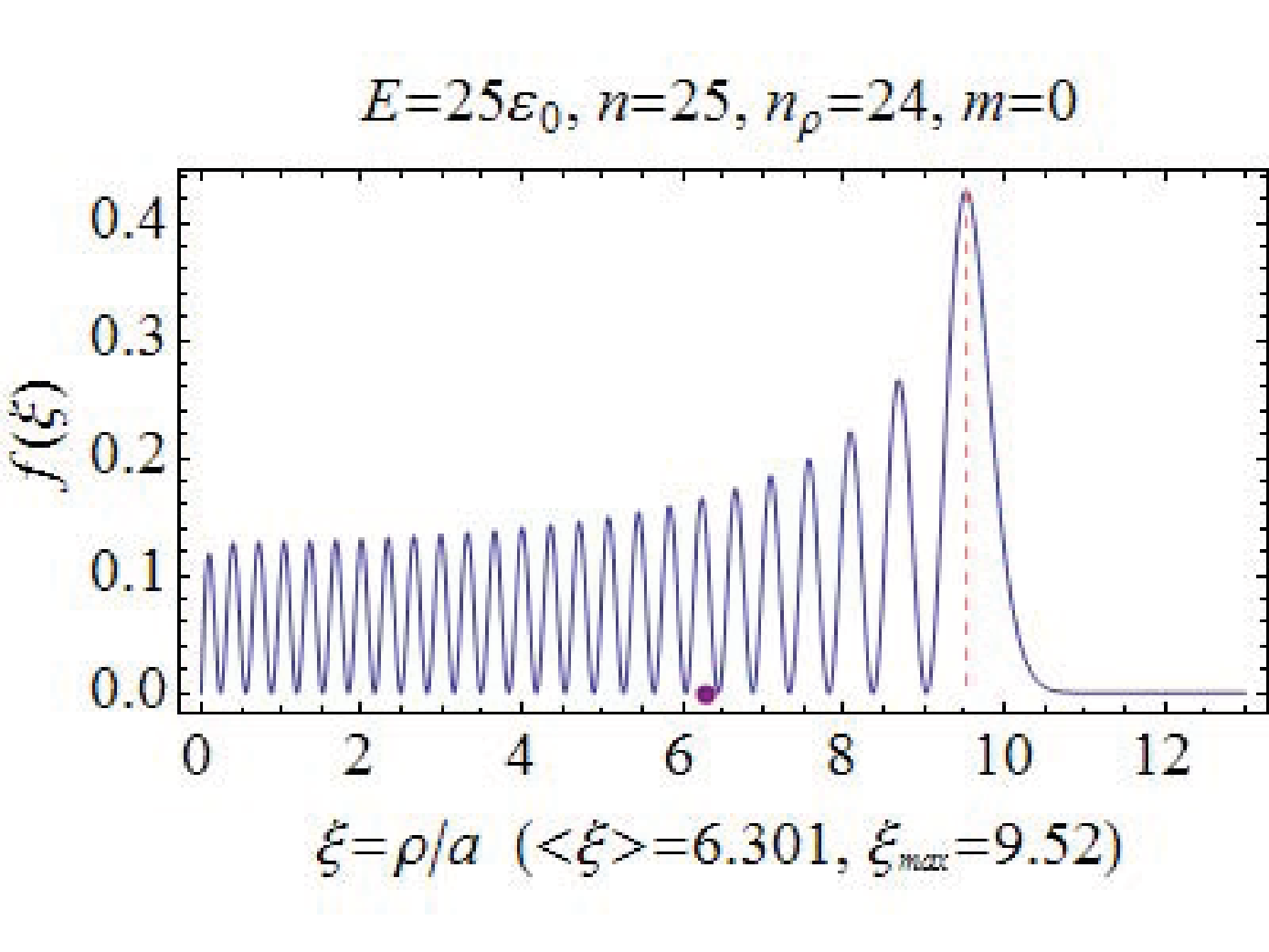}
\includegraphics[width=5.6cm,height=4.2cm]{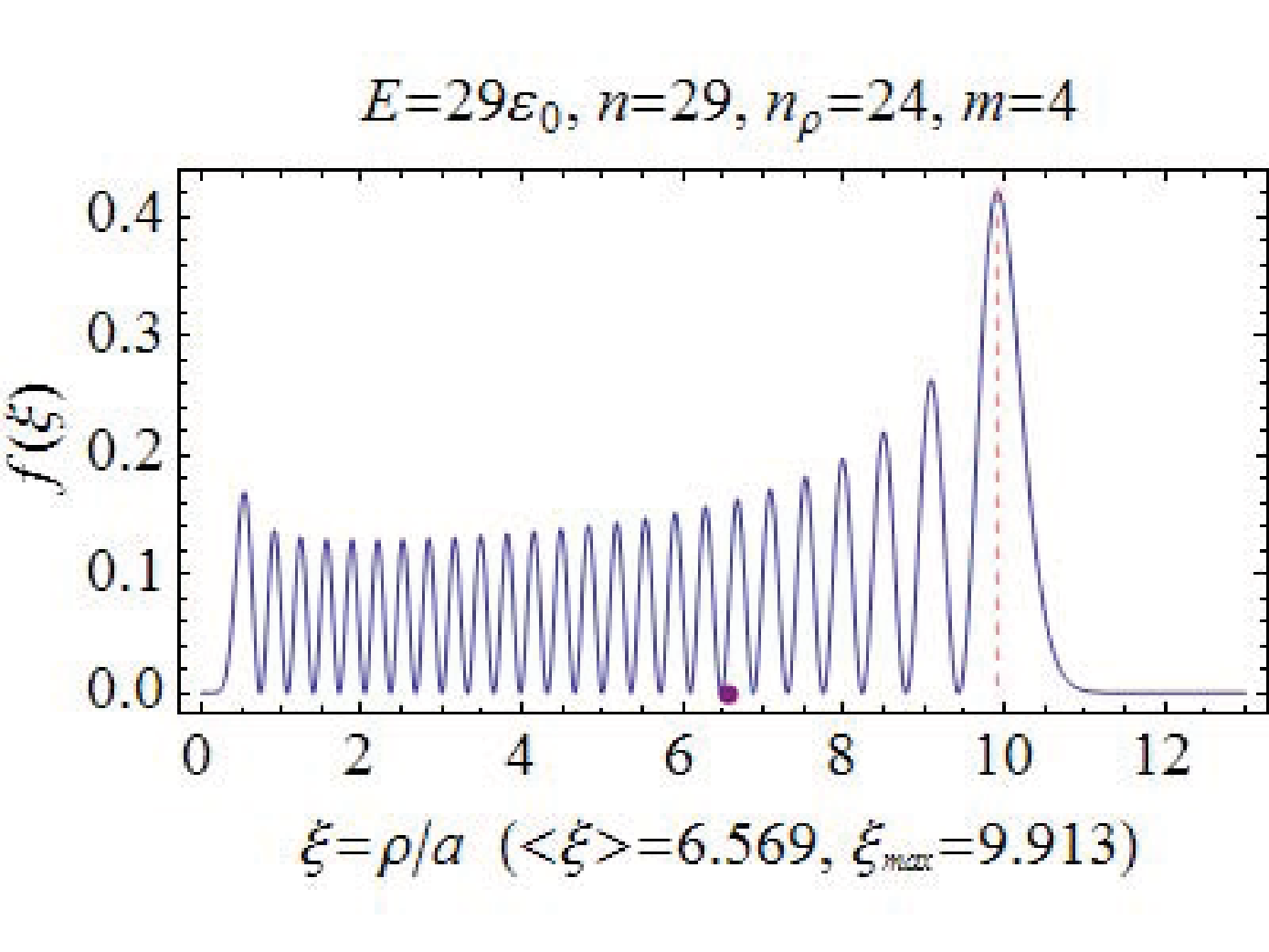}
\includegraphics[width=5.6cm,height=4.2cm]{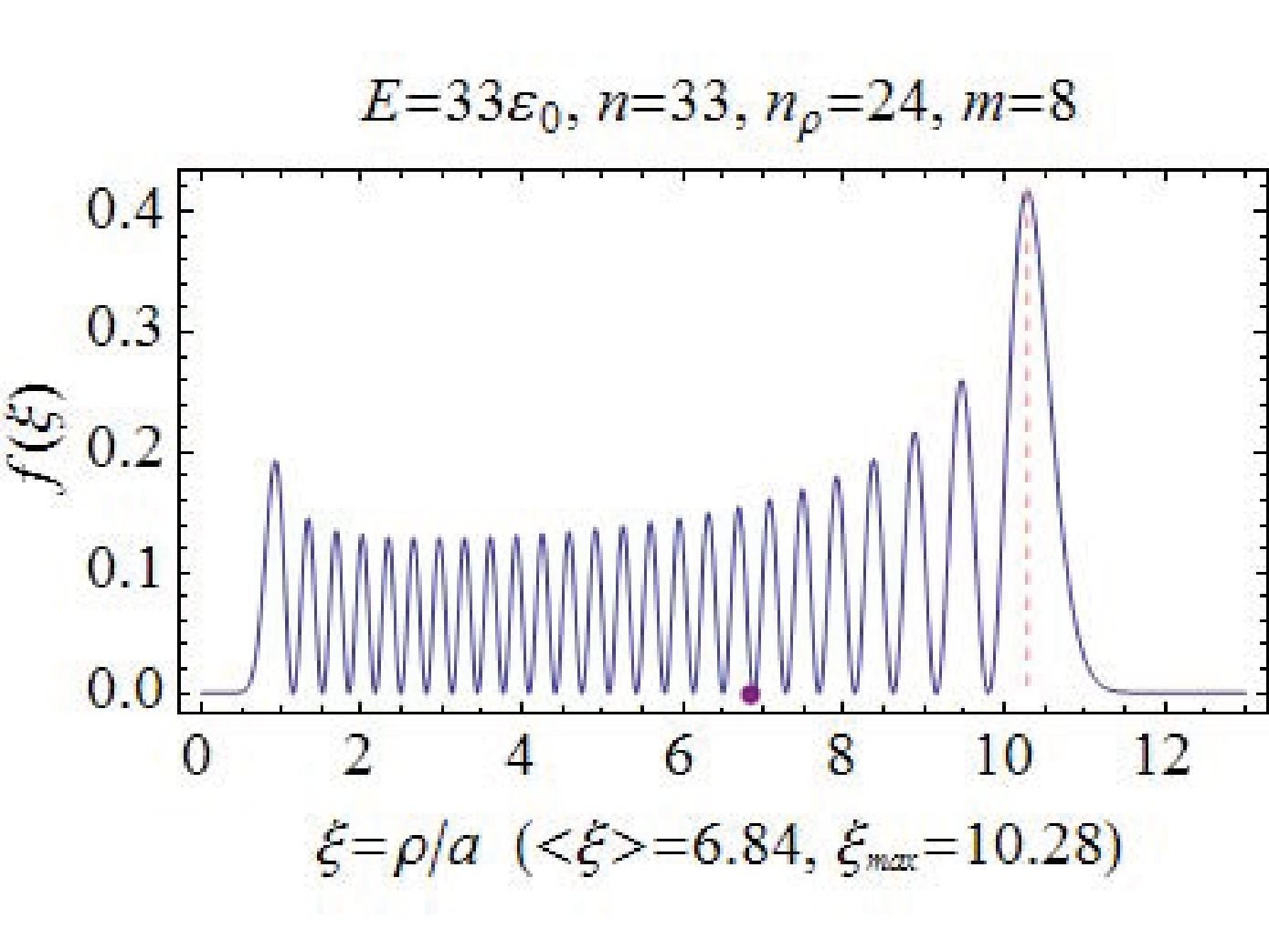}\\
\caption{Distribution curve of probability density of the electronic movement in Magnetic field }\label{}
\end{figure}

These graphs show electron's radial range is not related to its spin state but decided by both radial quantum number $ n_\rho $   and angular quantum number $| m|$.  Calculation shows that the  maximum probability density   located at  $\rho=\rho_{max}$, and it has the expression as,
\begin{equation}
\rho _{m\, a\, x}=\frac{1.1987+0.72156n_{\rho }+0.23395m}{1+0.03877n_{\rho }+0.00693m}a\approx \frac{1199+722n_{\rho }+234m}{1000+39n_{\rho }+7m}a
\end{equation}
where $0\leq 28, ~ 0\leq m\leq 12.$

From the Figure 1 some conclusions can be drew as follows:
(1)	The higher energy levels become, the wider the space range of electron motion will be and the fewer electrons will appear. This shows electrons have the preference for low-energy state.
(2)	The maximum of the electron probability density is only decided by radial quantum number. While, the maximum orbit radius of the probability increases obviously with radial quantum number, and it is decided by both radial quantum number and absolute value of angular quantum number m.
(3)	When $ m<0$, the electron energy levels are only decided by radial quantum number, and different m's probability remains same. However, the bigger the absolute value of m is, the wider the space range of electron motion will be. Then, the realization of this in physics will be more difficult, because when the real electron's space range is wider, more interaction between electron and lattice environment appears, thus non-interaction electron gas model will be more different from what really happens in crystal lattices. Therefore, electrons have the preference for small absolute value of angular momentum.
(4)  When electrons' physical surface density is bigger than specified Hall surface density, electrons will fill the higher energy level. Thus, electron gas Hall surface density will be weaker with decreasing external magnetic field. The weaker the magnetic field is, the higher levels electrons will fill and more interactions of high- energy electron with crystal lattice will happen. In this way, it will be more difficult for superfluidity and superconductivity to take place.

\section{Energy degeneracy  density of  the  electron gas in a magnetic field}
Eq.(14) shows that if the sum of quantum number is equal to  $n_\rho+|m|$, electron energy in the magnetic field also changes with spin state. The electron energy caused by a spin quantum number $\lambda=1 $ is $\epsilon_0$ higher than that when a spin quantum number $\lambda=-1 $,  Namely, spin causes electron motion's energy division or part of energy degeneracy's removal. When $m<0$ , Eq.(14)changes to
\begin{equation}
E_{n_{\rho }m\, \lambda }=\frac{p_z{}^2}{2\mu }+(n_{\rho }+\frac{\lambda +1}{2})\frac{e\hbar B}{\text{$\mu $c}},\, \text{   }n_{\rho }=0,1,2,\text{...};\, \, \lambda =-1,1
\end{equation}
This shows that when $m<0$ , the energy degeneracy is infinity (also shown in [19,21]). In other words, the energy degeneracy has nothing to do with angular quantum number.
If a great amount of electrons having momentum $p_z $ move to a magnetic field along z direction (assuming $p_z^2/2\mu$, compared with the interaction of the electron  and magnetic field,  is too small to be noticed) , two dimensional dilute electronic gas forms. In terms of energy degeneracy, when  $m<0$ , one energy level should be occupied by infinitely many electrons. But this is not what really happens. In fact, every electron has its own motion range. So, in a limited space the actual number of electrons at one energy level filling number is finite. For two dimensional dilute electronic gas, in one energy level electronic states' probability density   reads

\begin{equation}
2\pi \rho \left|\phi (\rho ,\varphi ,z,s)|^2\right.=2\frac{(n_{\rho }+|m|)!}{a\, n_{\rho }!}e^{-\xi ^2}\sum _{k=0}^{n_{\rho }} \sum _{j=0}^{n_{\rho }} \frac{(-1)^{k+j}\text{Binomial}[n_{\rho },k]\text{Binomial}[n_{\rho },j]}{(|m|+k)!(|m|+j)!}\xi ^{2k+2j+2|m|+1}
\end{equation}
in which  $\xi= \rho/a$. Therefore, the expectation value of electron motion's area is

\begin{equation}
\langle S\rangle _{n_{\rho }m}=\pi \int _0^{2\pi }d\varphi \int _0^{\infty }\rho ^3|\phi (\rho ,\varphi ,z,s)|^2d\rho =\left(2\text{\textit{$n_{\rho }$}}+|m|+1\right)\frac{\text{\textit{hc}}}{\text{\textit{eB}}}
\end{equation}
When the energy level is definite,  the electron state number is decided by $n_\rho, m$ and $\lambda$,   When the quantum Hall effect appears, $ m\leq 0$, and electrons are completely polarized, i.e $  \lambda=-1$, then the energy level has nothing to do with $m$. According to Pauli principle, angular momentum quantum number can be 0,-1,-2,¡­£¬-m, so on the area of $\langle S\rangle _{n_{\rho }m}$  the number of electron state is

\begin{equation}
N_B=|m|+1
\end{equation}
Thus, in the magnetic field move electron's energy degeneracy (also named electron gas Hall surface density) per unit area and Hall resistance are:
\begin{equation}
n_B=\frac{N_B}{\langle S\rangle _{n_{\rho }m}}=\frac{|m|+1}{\pi (2n_{\rho }+|m|+1)a^2}=\frac{|m|+1}{2n_{\rho }+|m|+1}\frac{eB}{hc},~\text{\textit{$\rho_{xy}=\frac{B}{n_Bec}=\frac{\left.2n_{\rho }+|m\right|+1}{|m|+1}\frac{h}{e^2}$}}
\end{equation}
In the ground state with $ m¡Ü0,n_\rho =0$, Eq.(23) changes to
\begin{equation}
n_B=\frac{N_B}{\langle S\rangle _{n_{\rho }m}}=\frac{eB}{hc},~~\text{\textit{$\rho_{xy}=\frac{B}{n_Bec}=\frac{h}{e^2}$}}
\end{equation}

The electron Orbital magnetic moment reverses the external magnetic field, so when angular quantum number $m>0$, with the maximum of finite energy degeneracy $ n +m+1+( +1)/2$,  we can define respectively electron gas Hall surface density and quantum Hall resistance as

\begin{equation}
\text{\textit{$\text{\textit{$n$}}_B$}}\text{\textit{$=$}}\frac{\text{\textit{$n_{\rho }$}}+\text{\textit{$m$}}+1+(\lambda +1)/2}{2\text{\textit{$n_{\rho }$}}+\text{\textit{$m$}}+1}\text{\textit{$\frac{eB}{hc}$}}\text{\textit{$,$}}~~\text{\textit{$\rho$}}_{\text{\textit{$xy$}}}=\frac{2\text{\textit{$n_{\rho }$}}+\text{\textit{$m$}}+1}{\text{\textit{$n_{\rho }$}}+\text{\textit{$m$}}+1+(\lambda +1)/2}\text{\textit{$\frac{h}{e^2}$}}
\end{equation}
Further detailed interpretation of unified formulation of the integer and the fractional quantum Hall effects can be found in our recent work\cite{wll}.

\section{Conclusion remarks }
In summary, in a strong magnetic field by solving the Pauli equation, single electron's energy levels and the wave functions are presented. In fact, this problem was  solved, under Landau gauge, by Landau in 1930. However, the wave function given in the Landau gauge is the wave function of one-dimensional harmonic oscillator whose infinite energy degeneracy can be realized from the  arbitrary choice of coordinates center  of the  harmonic oscillator. In addition it needed to suppose that  electron moves in a limited area, such that  momentum  can be quantized, and then   a quantum Hall resistance can be understand. In this paper, different from the previous studies, symmetric gauge is used and quantum Hall resistance is obtained without any suppose conditions. It is concluded that when  $m\leq 0$, energy levels has no relation to $m$   and energy level degeneracy is theoretically infinite. This is supposed to be the very cause of the quantum Hall effect. Further study on the problem will be provided elsewhere.

\section{Acknowledgments}

This work was supported by the National Natural Science Foundation of China (10447005, 10875035),
an open topic of the State Key Laboratory for Superlattices and Microstructures (CHJG200902), the
scientific research project in Shaanxi Province (2009K01-54) and  Natural Science Foundation of Zhejiang Province (Y6110470),


\begin{thebibliography}{10}

\bibitem{klizing1}
K. von Klitzing, G. Dorda, M. Pepper. "New Method for
High-Accuracy Determination of the Fine-Structure Constant Based on
Quantized Hall". Physical Review Letters {\bf 45}: 494¨C497 (1980)
\bibitem{klizing2}
  Klaus von Klitzing, Nobel Lecture: The quantized Hall effect, Rev. Mod. Phys. 58, 519-531 (1986).
\bibitem{tsui1}
D.C. Tsui, H.L. Stormer, A.C. Gossard . Two-Dimensional
Magnetotransport in the Extreme Quantum Limit. Physical Review
Letters 48: 155(1982)
\bibitem{laughlin1}
R.B. Laughlin. Anomalous Quantum Hall Effect: An
Incompressible Quantum Fluid with Fractionally Charged Excitations.
Physical Review Letters 50: 1395(1983)
\bibitem{stormer1}
H.L.Stormer, Nobel Lecture: The fractional quantum Hall effect, Rev. Mod. Phys. 71, 875-889 (1999); R.B.Laughlin, "Nobel Lecture: Fractional quantization", Rev.Mod. Phys. 71, 863-874 (1999).
\bibitem{klizing3}
Tapash Chakraborty and Klaus von Klitzing. Taking stock of the quantum Hall effects: Thirty years on. arxiv:1102.5250v1.
\bibitem{AFS}
  T. Ando, A.B. Fowler, and F. Stern, Electronic prop-erties of two-dimensional systems, Rev. Mod. Phys. 54,437-672 (1982).
\bibitem{halperin}
 B.I. Halperin, Statistics of quasiparticles and the hierar-chy of fractional quantized Hall states, Phys. Rev. Lett.52, 1583-1586 (1984).
 \bibitem{ASW}
 D. Arovas, J.R. Schrier, and F. Wilczek, Fractional statistics and the quantum Hall effect, Phys. Rev. Lett.53, 722-723 (1984).
 \bibitem{wll}
 Jianhua Wang, kang Li and Shuming Long, An alternative unified formulation for integer and fractional quantum Hall effects, arXiv:1107.0759
\bibitem{Long}
 Shuming Long,Jiewu Zhu and Yanqing Sun, Methods of mathematical physics ,China's Xian, Shaanxi people's publishing house,,97-153(2002)




\end{thebibliography}
\end{document}